\documentclass[aip,graphicx]{revtex4-1}

\usepackage{graphicx}
\usepackage{hyperref}
\usepackage{amsmath, amssymb, textcomp, gensymb}
\usepackage[scientific-notation=true]{siunitx}

\begin{document}

\title{Towards New Liquid Crystal Phases of DNA mesogens}

\author{Kit Gallagher}
\affiliation{Cavendish Laboratory, University of Cambridge, Cambridge CB3 0HE, United Kingdom}%
\author{Jiaming Yu}
\affiliation{Cavendish Laboratory, University of Cambridge, Cambridge CB3 0HE, United Kingdom}%
\author{David A. King}
\affiliation{Cavendish Laboratory, University of Cambridge, Cambridge CB3 0HE, United Kingdom}%
\author{Ren Liu}
\affiliation{Cavendish Laboratory, University of Cambridge, Cambridge CB3 0HE, United Kingdom}%
\author{Erika Eiser}
\email{erika.eiser@ntnu.no}
\altaffiliation{\\PoreLab, Department of Physics, Norwegian University of Science and Technology, N-7491 Trondheim, Norway}%
\affiliation{Cavendish Laboratory, University of Cambridge, Cambridge CB3 0HE, United Kingdom}%

\date{\today}

\begin{abstract}
Short, partially complementary, single-stranded (ss)DNA strands can form nanostructures with a wide variety of shapes and mechanical properties.  
It is well known that semiflexible, linear dsDNA can undergo an isotropic to nematic (IN) phase transition~\cite{strzelecka1988multiple,nakata2007end} and that sufficiently bent structures can form a biaxial nematic phase~\cite{Etxebarria2008,yang2018phase}.
Here we use numerical simulations to explore how the phase behaviour of linear DNA constructs changes as we tune the mechanical properties of the constituent DNA by changing the nucleotide sequence.
The  IN phase transition can be suppressed in so-called DNA ``nunchakus'': structures consisting of two rigid dsDNA arms, separated by a sufficiently flexible spacer.
In this paper, we use simulations to explore what phase behavior to expect for different linear DNA constructs.
To this end, we first performed numerical simulations exploring the structural properties of a number of different DNA oligonucleotides,  using the oxDNA package~\cite{ouldridge2011structural,OxDNA}.
We then used the structural information generated in the oxDNA simulations to construct more coarse-grained models of the rodlike, bent-core, and nunchaku DNA. 
These coarse-grained models were used to explore the phase behavior of suspensions of the various DNA constructs.
The approach explored in this paper makes it possible to ``design'' the phase behavior of DNA constructs by a suitable choice of the constituent nucleotide sequence. 
\end{abstract}

\maketitle 

\section{Introduction}
In his seminal 1949 paper~\cite{Onsager1949}, Onsager predicted that thin, rodlike colloids could undergo a purely entropy-driven transition from an isotropic liquid to a nematic liquid crystal.

Entropy-driven phase transitions of suspensions of  non-spherical, hard colloidal particles have been the subject of many experimental, theoretical and numerical studies, see e.g.  \cite{Anderson2002, Eppenga1984,engel2015computational,dij151}. 
In addition, many experimental and simulation studies focused on thermotropic molecular liquid-crystal formers, amongst which `bent-core' or `banana-shaped' molecules garnered particular interest because they display a rich phase behaviour that includes different smectic phases. 
Such liquid crystals are of interest in view of their potential applications in non-linear optics and display technology \cite{Takezoe2006, Etxebarria2008,reddy2006bent,chi191,chi211,kubala2021silico}. 
It is challenging to synthesize colloidal bent-core mesogense with sufficiently narrow size- and angle-distributions such that a clear liquid crystal (LC) phase transition could be observed \cite{Yang2018}. 
An important advance was made by Ferna\'ndez-Rico et al.\cite{fer201}, who succeeded  in synthesizing banana-shaped colloids with a low polydispersity, for which the bending-angle could be controlled continuously. 
Ferna\'ndez-Rico et al.\cite{fer201} performed confocal-microscopy studies on suspensions of their banana-shaped colloids to explore the various LC-phases directly as function of the bending angle. 
Their observations could be reproduced in variational mean-field theory and Monte Carlo simulations\cite{chi211}.

An attractive approach to design perfectly  monodisperse, bent-core systems with no polarity or attractive inter-particle potentials is the use of relatively short, single-stranded DNA chains.
Such chains had earlier been used to create self-assembling DNA-nanostars~\cite{seeman1982nucleic,biffi2013phase,xing2019structural} and a wide range of DNA-origami structures~\cite{rothemund2006folding,liu2014dna-i-motif-reveiw,Cha2015,Wang2018,seeman2017dna,canary2022nadrian}. 
The large difference in persistence length of double- and single-stranded DNA has also been used to explore linear, double-stranded (ds)DNA as liquid-crystal former\cite{livolant1996condensed,fraccia2016liquid,Salamonczyk2016,de2016hierarchical,winogradoff2021chiral}. 
However, in experiments, a key challenge for using DNA lies in the design of thermodynamically stable DNA sequences that can then organize into the desired LC-phase.
A practical problem is that designed DNA sequences tend to be relatively short (for cost reasons) and that, for the typical aspect ratios of such short DNA-constructs, high concentrations are needed to reach the IN-phase transition.  
As a consequence, it is costly to prepare many different mesogenic DNA-constructs. 

Salamonczyk et al.\cite{Salamonczyk2016} introduced a tailor-made, coarse-grained model to specifically emulate the smectic phase transition of DNA nunchakus they studied in experiments. 
There they used a 20 Tymine long spacer to connect two rigid dsDNA duplexes, which were represented by hard cylinders connected by two beads via appropriate springs. This choice allowed the two rigid arms to align parallel to each other thus promoting smectic alignment. However, this model did not reflect the dependence of the bending angle between the two arms as function of ssDNA linker-length. Here we show how to use oxDNA\cite{ouldridge2011structural,OxDNA}, a semi-coarse-grained simulation model that provides structural and thermodynamic properties of any DNA construct, to design nearly rigid rods, and corresponding rigid, bend core and flexible ‘nunchakus’ mesogens, and then test these in the laboratory. Subsequently, we transposed our computationally expensive oxDNA model into a further coarse-grained bead-spring model, developed by Xing et al. \cite{xing2019structural} for DNA nanostars, using LAMMPS\cite{LAMMPS}. This approach allowed simulating larger systems of DNA-mesogens, and provides a better predictor of the experimental phase behaviour of such systems. We identified individual phases by measuring the diffusion coefficients of the mesogens along the different coordinates. While literature exists on the dynamic behaviour of mesogens within liquid crystal phases \cite{deMiguel1991gayberne, rey1997diffusion}, this method was not used previously to identify the nature of the smectic phase.

\section{Simulation Model}

\subsection{Mapping oxDNA findings to a Coarse-Grained Bead-Spring Model} \label{sec:SimMolecules}

We base our coarse-grained bead-spring model\cite{xing2019structural} of rigid rods and nunchakus with different flexibility on the molecular structure and interaction parameters obtained from the semi-empirical, more detailed oxDNA calculations\cite{doye2013coarse, ouldridge2011structural} of DNA sequences, which we tested experimentally. The specific sequence and detailed experimental and oxDNA modelling are given in the supporting information.

\begin{figure}[ht]
  \centering
  \includegraphics[height=11.8cm]{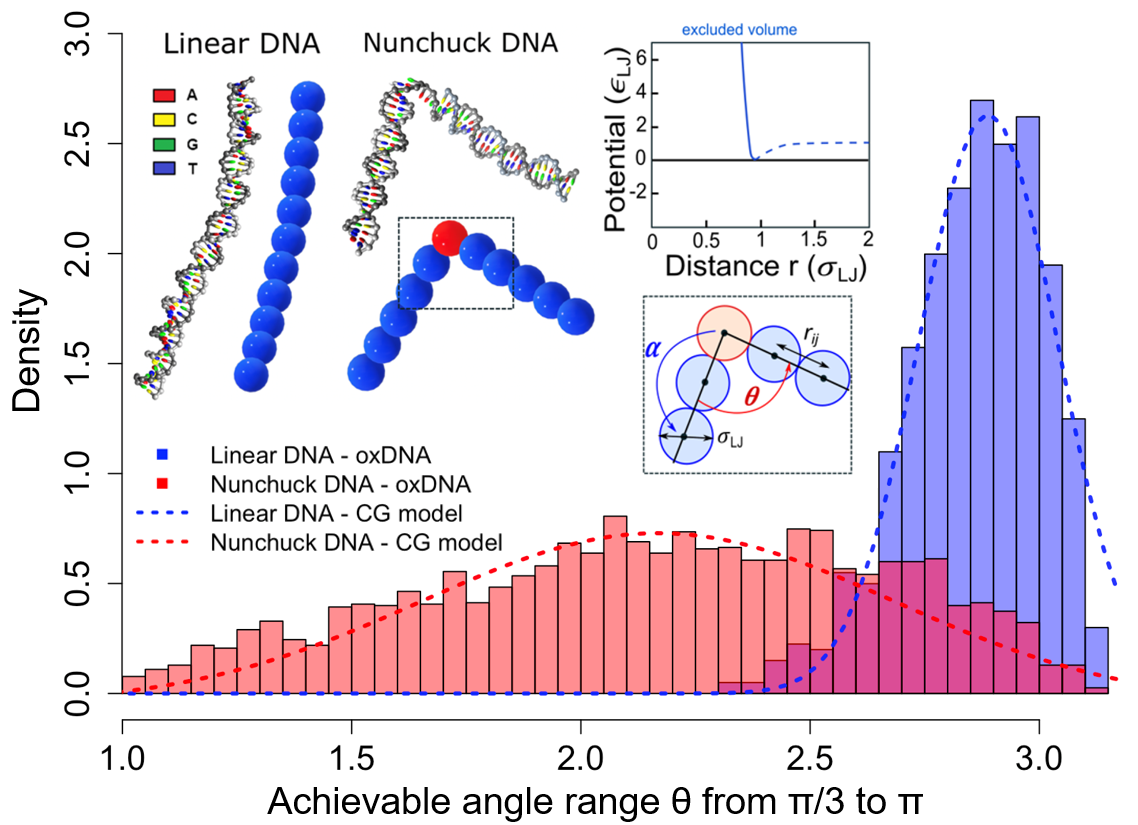}
  \caption{Both schematic representations of our coarse-grained bead-spring model and the oxDNA calculation below the melting temperature of the DNA sequences used.}
  \label{fgr:nunchaku_model}
\end{figure}

The nunchaku molecules are formed of two rigid rods connected by a flexible linker, as depicted in Figure \ref{fgr:nunchaku_model}. Each rod is modelled as a series of beads that have a diameter of \SI{2}{\nano\metre}, corresponding to approximately 6 base pairs of ds-DNA \cite{Arnott1972}. These beads are connected by harmonic springs, and kept linear by an angular potential with a minimum at $180\degree$. All molecules considered here are shorter than the persistence length of double-stranded DNA, which is $\sim$ 50\;nm \cite{Garcia2007}. So the approximation of near-perfect rigidity (and linear rods) is valid, and justified further in the SI.

Two separate cases for the central angular potential connecting the two rods were considered, allowing for bent-core and flexible mesogens. For bent-core molecules, the minimum of the angular potential was set to $\theta = 150\degree$ allowing for small fluctuations around $\theta$, thus giving all mesogens approximately the same opening angle. In the flexible case, the energy scale of the angular potential is reduced while the minimum is kept at at $180\degree$, so that the opening angle of the mesogens can vary considerably. This choice was based on our oxDNA analysis of the flexibility provided by single-stranded DNA that links the two arms in the nunchakus. These oxDNA simulations, detailed in the SI and stated also in previous work\cite{xing2019structural}, suggest that an ssDNA linker of minimum length of 6 bases (roughly one bead size) provides a wide probability distribution of $\theta$ as shown in Figure\,\ref{fgr:nunchaku_model}. In the same figure we also show the angle-probability distribution for the same sequence with no ssDNA flexibility - this quasi-hard rod system was used to calibrate our simulations with regard to isotropic to nematic phase transitions as function of aspect ratio.

The model was implemented in LAMMPS\cite{LAMMPS}, using reduced units, based on the fundamental units of mass $m$, Lennard-Jones (LJ) energy-scale $\epsilon_{LJ}$, LJ length-scale $\sigma_{LJ}$ and the Boltzmann constant $k_{B}$. These also define a characteristic time scale $\tau = (m\sigma_{LJ}^{2}/\epsilon_{LJ})^{1/2}$. The LJ quantities are defined by the form of the inter-molecular interaction potential, for which a shifted, cut-off potential (potted in Figure\,\ref{fgr:nunchaku_model}) is used:

\begin{equation} \label{eq:lj_potential}
U_{ij} = 4\epsilon_{LJ} \left[ \left( \frac{\sigma_{LJ}}{r_{ij}} \right) ^{12} - \left( \frac{\sigma_{LJ}}{r_{ij}} \right) ^{6}	\right] + \epsilon_{LJ} \qquad	 r_{ij} < r_{c} = 2^{1/6} \sigma_{LJ}.
\end{equation}
In this Weeks\textendash Chandler\textendash Andersen (WCA) style potential a cut-off distance $r_{c} = 2^{1/6} \sigma_{LJ}$ was chosen to represent the purely repulsive excluded volume interactions between the beads, ensuring that all phase transitions observed are truly entropy\textendash driven. Neighbouring beads are connected by the harmonic potential:

\begin{equation} \label{eq:harmonic_bond}
V_{bond} = K_{bond}(r-r_{0})^{2}
\end{equation} where $r_{0}$ is the equilibrium bond distance and $K_{bond}$ is the stiffness of the harmonic bond. We set $r_{0}$ to $0.96\sigma_{LJ}$ and $K_{bond}$ to $500\epsilon_{LJ}/\sigma_{LJ}^{2}$ throughout, allowing for minimal disturbances around the equilibrium distance. As the beads themselves have a radius $0.5r_{c}$ (equivalent to $0.56 \sigma_{LJ}$), this gives a small overlap between neighbouring beads. The angular potential is given by:

\begin{equation} \label{eq:harmonic_angle}
V_{angle} = K_{angle}(\theta-\theta_{0})^{2}
\end{equation} where $\theta_{0}$ is the equilibrium bond angle and $K_{angle}$ is the spring constant that controls the allowed distribution of the angle. We set $\theta_{0}^{rod} = 180\degree$ within each rod and $\theta_{0}^{bend} = 150\degree$ for the bent-core molecules. We chose $K_{angle} = 500\epsilon_{LJ}/rad^{2}$ throughout, except for the central angle in the flexible nunchakus, where it is given as $0.1\epsilon_{LJ}/rad^{2}$. This corresponds to the ss-DNA, represented by a separate bead in the centre of the molecule, with differing mechanical properties.

\subsection{Simulation Details} 

All molecular dynamics (MD) simulations presented here were performed using Langevin dynamics in LAMMPS \cite{LAMMPS}.Simulations were conducted on a system of 1000 particles, with a time step of $0.005\tau$, unless otherwise stated. Further, the simulations were performed within an oblong box defined by the Cartesian axes, with periodic boundary conditions \cite{Frenkel2002}. The aspect ratio of this box may be varied, to support phase formation for anisotropic mesogens and evaluate correlation functions over longer lengths without increasing the size of the system.

The system was initially configured in a dilute, isotropic state, prepared by random placement and orientation of the mesogens in a simulation box, using a Monte Carlo algorithm. Any molecules that did not fit in the simulation box or those that overlapped with the existing units were discarded, and the placement procedure was repeated until all particles were added. When studying high volume fractions, simulations were initiated from a perfectly ordered square crystalline phase, with all molecules aligned along a common axis. Care was taken to ensure stability of this ordered phase, by confirming molecules did not overlap and that the internal energy was conserved after equilibration.

An isenthalpic ensemble was used to expand/contract the size of the simulation box, hence varying the volume fraction of the mesogen system, while a microcanonical $(N,V,E)$ ensemble was used to thermally equillibrate the system at each volume fraction studied. Time integration was evaluated using the Nos\'e\textendash Hoover thermostat\cite{Nos1984, Hoover1985}, natively implemented in LAMMPS \cite{Shinoda2004}, with a damping time $\tau$.

A usual simulation run consisted of multiple stages, alternating between these two ensembles: Contraction stages varied the volume fraction and equillibration stages. Typically \num{2e4} steps (each of duration $0.005\tau$) were simulated in the isenthalpic ensemble, followed by \num{2e6} steps in the microcanonical ensemble, to allow the system to reach equilibrium at this volume fraction. After equilibriation we sampled the system properties. To ensure the stability of the system, a Langevin thermostat \cite{Schneider1978} (also with a damping time of $\tau$) was used throughout, maintaining a temperature of $0.5\epsilon_{LJ}/k_{B}$, and energy conservation was verified over a range of timescales.

\section{Phase Characterisation}
Our linear, rod-like DNA mesogens display traditional liquid-crystal phase transitions with characteristic nematic and smectic order parameters. The nematic order parameter $S_{n}$ is given by:

\begin{equation} \label{eq:nem_order_param}
S_{n} = \left\langle P_{2}(\cos \varphi) \right\rangle = \left\langle \frac{3}{2}\cos^{2}\varphi - \frac{1}{2} \right\rangle 
\end{equation} where $P_{2}$ denotes the second Legendre polynomial \cite{DeGennes1993}. This is non-trivial to calculate when the system director is not known (i.e. in the absence of an external field). Therefore, we used the approach taken by Frenkel et al. \cite{Frenkel1985} - further details are provided in the SI.

Similarly, the smectic order parameter is defined as

\begin{equation} \label{eq:smec_order_param}
S_{s} = \frac{1}{N} \left\lvert \sum_{j=1}^{N} \exp \left( {\frac{2\pi}{L}i\textbf{r}_{j} \cdot \boldsymbol{\hat{n}}} \right) \right\rvert
\end{equation} for layers of periodicity $L$ perpendicular to the nematic director $\boldsymbol{\hat{n}}$; $\textbf{r}_{j}$, denotes the centre\textendash of\textendash mass position of the $j$\textendash the molecule \cite{Dussi2018}.

It is also instructive to introduce a length-dependent orientational order parameter  to verify that this order is truly long-ranged, and not simply due to short-range steric effects. We consider an $l$-th rank, pair-wise correlation function, where $g_{l}(r)$ gives the correlation between the orientation of two particles separated by a distance $r$:

\begin{equation} \label{eq:PairWise_correlation}
g_{l}(r) = \frac{\langle P_{l}(\boldsymbol{\hat{u}_{i}}\cdot \boldsymbol{\hat{u}_{j}}) \delta(r_{ij}-r)\rangle}{\langle  \delta(r_{ij}-r) \rangle}
\end{equation} where $\boldsymbol{\hat{u}_{i}}$ is the director for molecule $i$, and $r_{ij}$ is the separation between a given pair of molecules $i$ and $j$ \cite{Zannoni1979}. In the disordered phase, this pair-correlation function decays to zero, while in the ordered phase it decays to the square of the orientational order parameter \cite{Frenkel1985b}:

\begin{equation} \label{eq:PairWise_simplification}
\lim_{r \to \infty}g_{l}(r) = \left\langle P_{l} \right\rangle ^{2}
\end{equation}

It is worth noting here that the maximum separation between particles is half the size of the simulation region, due to the periodic boundary conditions \cite{Frenkel1985c}, and that this will vary over the duration of the simulation. To determine order over longer length scales without increasing the volume of the system, we varied the aspect ratio of the simulation region, and sampled correlations along the long axis of the box.

The explicit calculation of $g_{l}(r)$ scales as $\mathcal{O}(N^{2})$, and requires the storage of all necessary angles, limiting the resolution possible \cite{Soper1998}. We may avoid this problem by expanding the function in terms of spherical harmonics, and sum the contributions for all pairs to a given molecule. Further details of this approach are provided in the SI. 

\newpage
\section{Results and Discussion}

\subsection{Rigid Rod Order Parameters}
A system of rigid rods was used to benchmark the analysis methods in comparison to theoretical predictions derived by Onsager \cite{Onsager1949}, and in particular the isotropic\textendash nematic phase transition which has been verified computationally through both Monte Carlo \cite{Frenkel1984, Lee1987} and molecular dynamics simulations \cite{Allen1987, Camp1996}, as well as experimentally \cite{Kubo1979,Oldenbourg1988,Fraden1993}.

\begin{figure*}
 \centering
 \includegraphics[height=12cm]{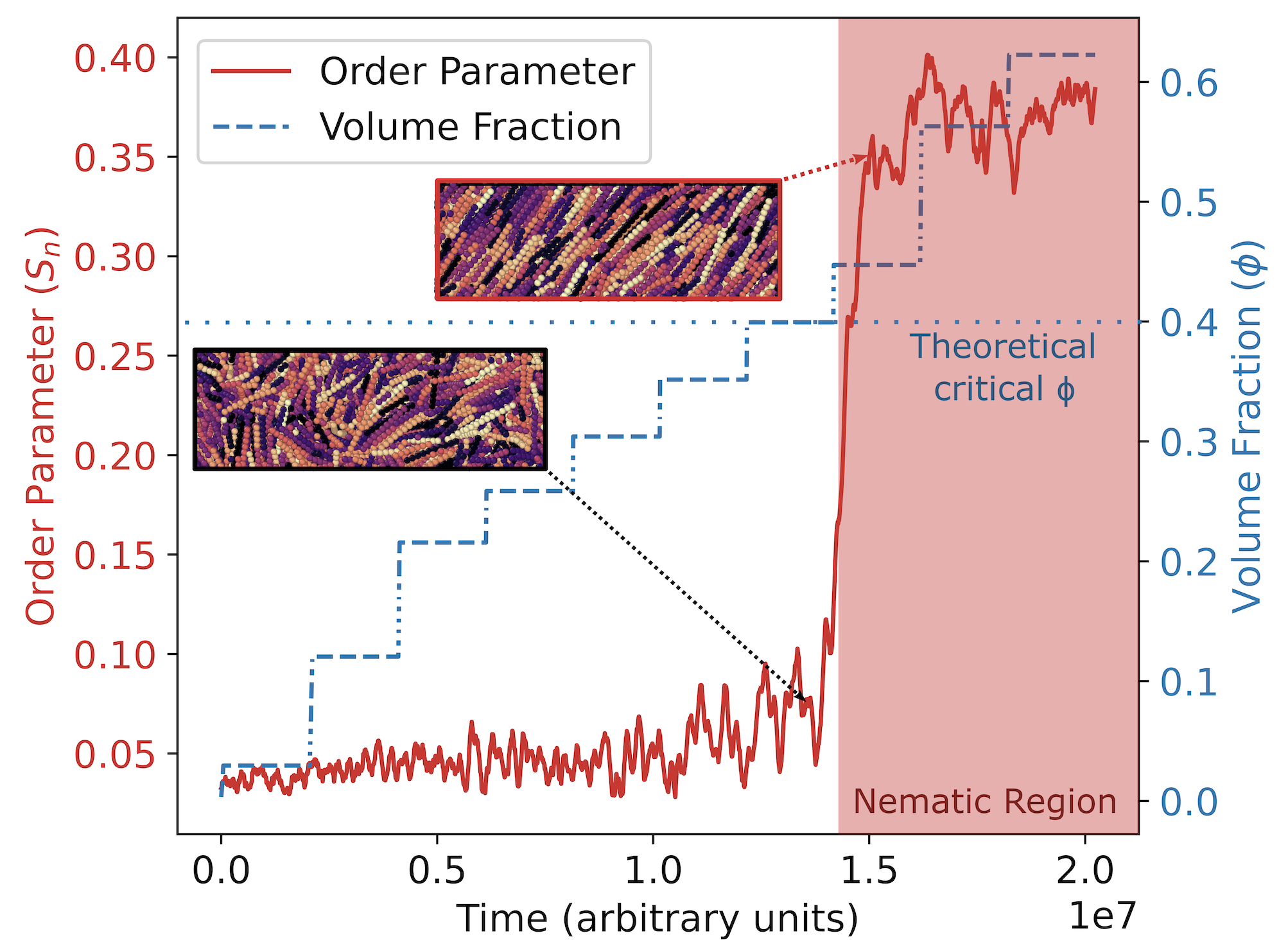}
 \caption{The evolution of the volume fraction and the nematic order parameter over the timescale of the simulation, for a system of 1000 rigid rods with aspect ratio 10. The discrete change in $S_n$ (red line) occurs at about $\phi \sim 0.4$, denoting the transition to the nematic phase, with the system depicted explicitly on either side of this transition. Note that the contraction steps, where the volume fraction is increased, are not of equal duration, and so do not correspond to equal changes in the system volume; rather they are chosen to highlight the phase transition. }
 \label{fgr:rr_nemorderparam}
\end{figure*}

For rigid rods with an aspect ratio $L/D = 10$, we confined the possible critical volume fraction for this lyotropic transition to the range $0.39 < \phi< 0.44$, as observed in Figure \ref{fgr:rr_nemorderparam}, achieving good agreement with Onsager's prediction of of $\phi  = 0.4$. This analysis was also repeated for longer rods with an aspect ratio of 16, with simulations identifying the critical volume fraction in the range $0.23 < \phi< 0.26$, in good agreement with the predicted value of $\phi  = 0.25$.

We also consider the phase behaviour upon expansion from a perfectly ordered, crystalline state. This has two advantages: it allows access to higher volume fractions that are not easily accessible through the shrinking procedure, and provides verification of the phase transitions previously observed. Ensuring that a novel phase is in true equilibrium has long been a challenge for liquid crystal simulators; however, non-equilibrium effects will manifest themselves in a hysteresis of the phase transition (variation in the critical volume fraction dependent on the direction of the transition) and can, therefore, be identified through this method.

The isotropic phase formation was observed in the region $0.38 < \phi< 0.41$, in good agreement with the previous simulations, confirming the equilibrium nature of this phase transition. The higher volume fractions accessed at the start of the simulation also suggested the existence of a smectic phase, with a sharp change in the smectic order parameter around $\phi  = 0.6$, however this has not been investigated further due to the wealth of literature on this already \cite{Frenkel1988, McGrother1996}.

\newpage
\subsection{Nunchaku Correlation Functions} \label{sec:nunchaku_Corr}

These techniques were subsequently applied to the nunchaku molecules introduced in Section \ref{sec:SimMolecules}. We consider first the flexible linker case, which we have verified provides an accurate coarse-grained model of the DNA-nunchakus using oxDNA \cite{OxDNA}. The simulation molecules were lengthened to include 15 beads instead of 10, as minimal order was observed at the lower rod aspect ratios.  

The linker flexibility was sufficient to ensure a quasi-isotropic initial angle distribution in the range $60\degree < \theta < 180\degree$, with smaller angles excluded due to steric repulsion between the rods. Unlike previous studies \cite{Salamonczyk2016}, this corrected the accurate possibility of the `fully-bent' ($\theta = 0\degree$) conformation where both rods are parallel, in agreement with the prior oxDNA results. The angle distribution we observed from the detailed oxDNA model indicated the low possibility of the fully-bent situation. More information from the oxDNA measurement can be found in SI.

\begin{figure*}
 \centering
 \includegraphics[height=12cm]{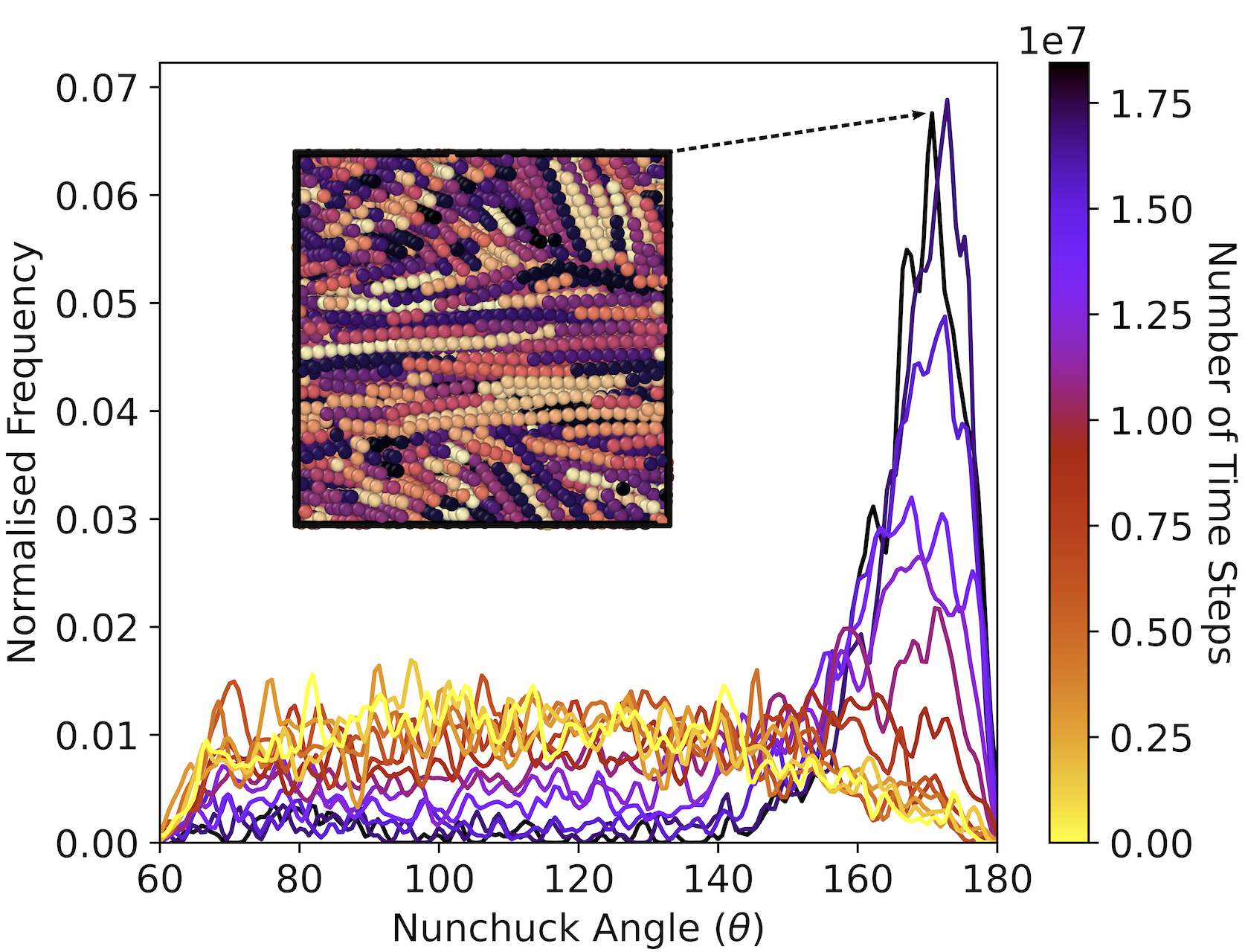}
 \caption{The kernel density estimate of the opening angle distribution as the volume fraction is reduced, for completely flexible nunchakus. Plotted for a system of \num{1e3} particles, with the distribution sampled every \num{1.5e6} time steps. Note the formation of a preferential angle at late times, corresponding to the formation of an ordered phase at high volume fractions, the final form of which is depicted in the inset.}
 \label{fgr:nun_angle}
\end{figure*}

Despite this flexibility, a strong preference for the linear configuration ($\theta = 180\degree$), was observed, with a quasi-nematic phase ($S_{n} = 0.6$) formed at volume fractions $\phi > 0.4$. However, no uniform, system-wide director was observed, with the preferential orientation varying across the simulation region shown in Figure \ref{fgr:nun_angle}. It is possible that periodic variation of the director occurs on a length scale greater than the size of the simulation region, but the computational resources required to simulate larger systems exceeded those available, preventing further characterisation of this phase.

To force the formation of more exotic LC phases with anisotropic (non-linear) mesogens, we also considered bent core rods, with a fixed opening angle. We considered opening angles $\theta = 60\degree, 90\degree, 120\degree, 150\degree, 165\degree$, and found that larger angles tend to form conventional nematic structures, while smaller angles are not conducive to ordered phase formation; in this section we focus on $\theta = 150\degree$.

Testing phase formation over a wide range of rigid rod lengths and opening angle values, we found that larger opening angles favoured nematic-like phase formation (with a maximal order parameter $S_{n} = 0.62$), but smaller angles did not form obvious alternate ordered phases. Using a pair-wise orientational correlation function (the calculation of which is discussed further in the SI), we could confirm that the order observed in this fixed angle system is indeed long-range, with a non-decaying correlation at long length scales.

\begin{figure*}
 \centering
 \includegraphics[height=12cm]{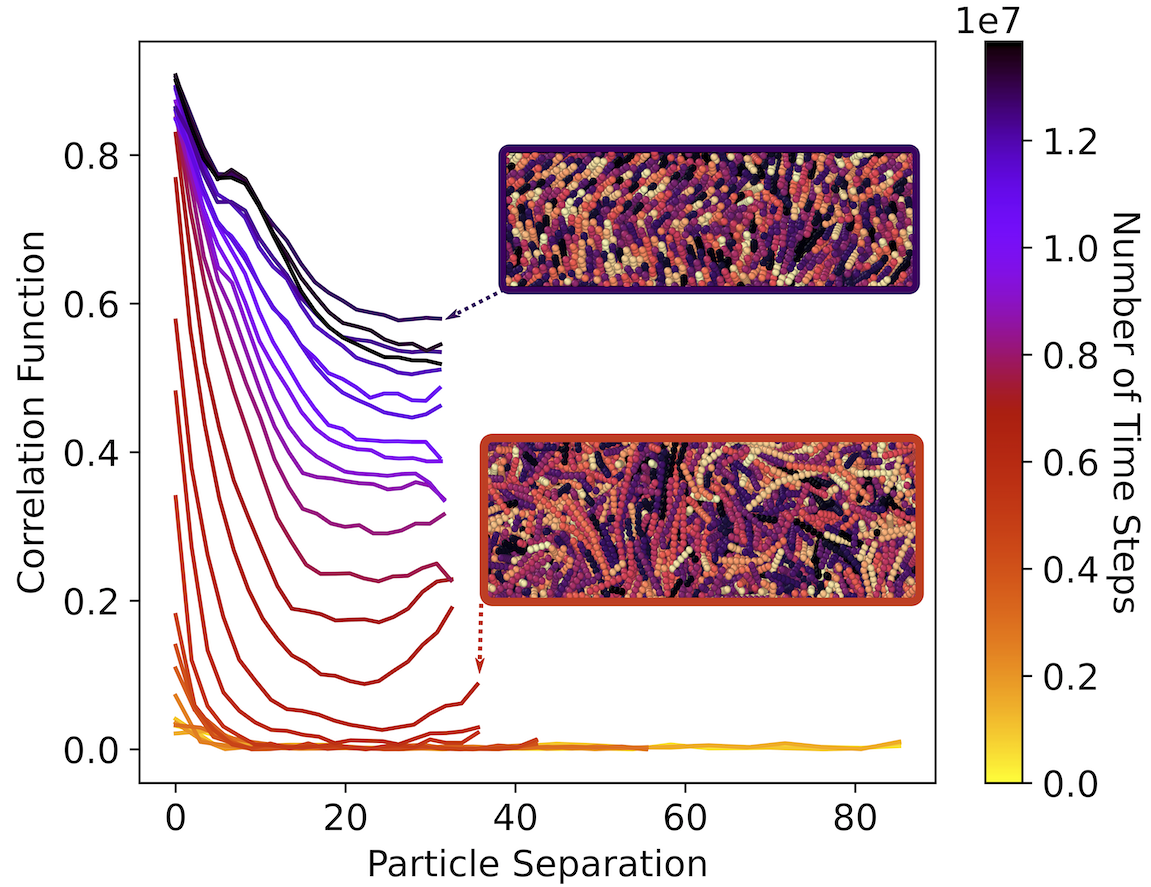}
 \caption{Orientational correlation function over time (as the volume fraction is reduced), for a system of \num{1000} nunchaku molecules, with a fixed opening angle of \SI{150}{\deg} and sampled every \num{7e5} time steps. Note the formation of an ordered phase at high volume fractions (late times), with sustained long-range order (i.e. no decay in the orientational correlation function). The maximum value of particle separation is determined by the size of the simulation box, and so shrinks over time.}
 \label{fgr:nun_cor}
\end{figure*}

This method is limited by the use of a single vector along the molecular axis to define the orientation of the molecule; two vectors are required to uniquely specify the orientation of a single nunchaku molecule. Using the molecular axis alone only accounts for quasi-nematic order and does not consider any biaxial or twisted nematic substructure. We therefore also considered additional orientational correlation functions for the bisector of each molecule, and normal vector to the plane of the nunchaku. The first Legendre polynomial is used to detect changes in sign of the bisector direction, as would be expected in a herringbone structure.

However, no statistically significant periodic components in the correlation function were observed over the length scale of the simulation region. A visual inspection of the inset in Figure \ref{fgr:nun_cor} suggests that the length of the simulation region is approximately half the full period of the repeating `twist' in the ordered phase, and so further research on larger simulation regions is required to verify these phases.

\newpage
\subsection{Dynamic Behaviour}
While the phase behaviour of rigid rods is well studied, much less is known about the dynamic properties of the individual mesogens within these quasi-ordered phases. Nevertheless, dynamic studies through MD simulations have enjoyed recent popularity in both computational and experimental work \cite{GayBalmaz2013, Zhao2013, Rey2013}. Here, we study the dynamic properties of the nunchaku system, and demonstrate the application of dynamic properties to static phase identification. In particular, we focus on the diffusion coefficient $D$, defined by the `power-law' equation for mean-squared displacement (MSD) as a function of time $t$ in $n$ dimensions\cite{Ernst2013}: 

\begin{equation}
\left\langle \left(x(t) - x_{0}\right)^{2} \right\rangle = 2nD_{\alpha}t^{\alpha}
\end{equation} which reduces to the pure-diffusive case when $\alpha=1$. Otherwise the process is characterised as subdiffusive ($\alpha < 1$) or superdiffusive ($\alpha > 1$) \cite{Metzler2000}. 

Displacements were only sampled over equilibration (constant volume) stages of our simulation, and the effect of the periodic boundary conditions was explicitly accounted for by offsetting additional displacements from boundary crossings. Linear regression analysis was used to predict the value of $\alpha$ for each run over a variety of volume fractions in the dilute limit. where each particle may exist in a non-overlapping free volume of rotation (a sphere circumscribed around the molecule). For rigid rods with an aspect ratio of 10, this occurs at $\phi_{d} = 0.015$. 

We found an average power of $ \alpha = 0.97 \pm 0.03 $ for the nunchaku molecules, and $ \alpha = 1.00 \pm 0.04 $ for the rigid rods, as expected for diffusive behaviour. 

\begin{figure*}
 \centering
 \includegraphics[height=6.8cm]{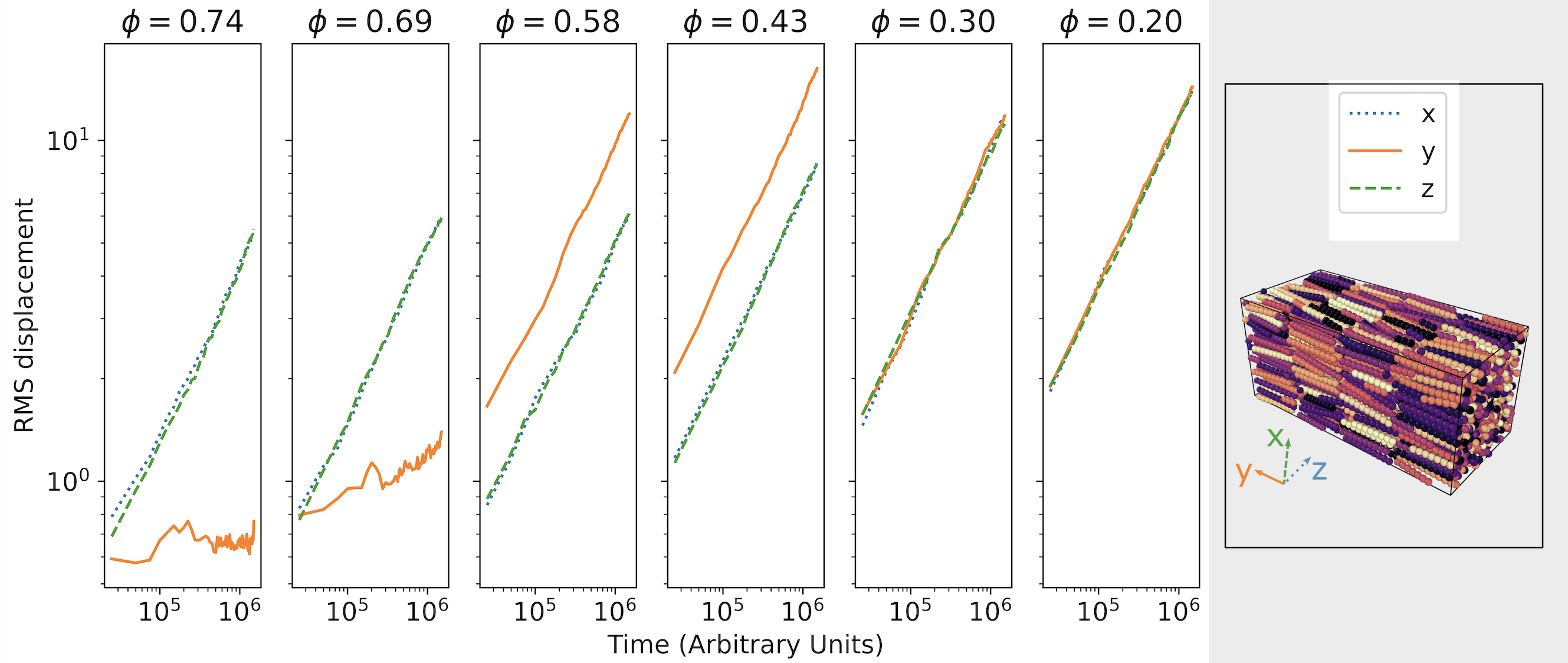}
 \caption{Root mean-squared displacements against time in a system of 1000 rigid rods with an aspect ratio 10, for a range of volume fractions $\phi$. The leftmost pair of plots correspond to the smectic phase, with restricted motion between $x-z$ layers along the $y$-axis. The centre pair give the nematic phase, with preferential displacement along the $y$-axis. Finally, the rightmost pair give the isotropic phase, with isotropic diffusion and no preferential direction. Note that no significant differences appear between the $x$- and $z$-directions in any phase, as these directions are equivalent in the phase structure.}
 \label{fgr:rigid_rms}
\end{figure*}

As the volume fraction was increased, nunchakus tend towards subdiffusive behaviour, with a reduction in the value of $\alpha$ to $0.12 \pm 0.02 $ at $\phi = 0.69$. We also observe anisotropy in the coordinate-wise displacement in non-isotropic, ordered phases, as demonstrated in Figure \ref{fgr:rigid_rms}, which considers three different phase regimes for a system of rigid rods. Here the Cartesian axes are used, as simulations are configured in a crystalline phase with molecules orientated along the $y$ axis, and smectic layers forming in the $x-z$ plane. Phase transitions for this system were previously identified in the regions $0.58<\phi<0.62$ for the smectic\textendash nematic, and $0.38<\phi<0.43$ for the nematic\textendash isotropic transition. 

At the highest volume fractions ($\phi \geq 0.62$), the diffusion along the $y$ axis is significantly reduced, due to the restricted motion between layers in the smectic phase. Below this, a nematic phase is observed with increased diffusivity along the molecule's director, corresponding to an increased intercept in logarithmic space. The degree of dynamic anisotropy here, quantified by the ratio of diffusion coefficients $D_{y}/D_{x}$, is $2.12$ for $\phi = 0.58$. At the lowest volume fractions ($\phi \leq 0.38$), there is no preferred direction in the system and motion is isotropic and truly diffusive (as $\alpha = 1$). 

\begin{figure*}
 \centering
 \includegraphics[height=12cm]{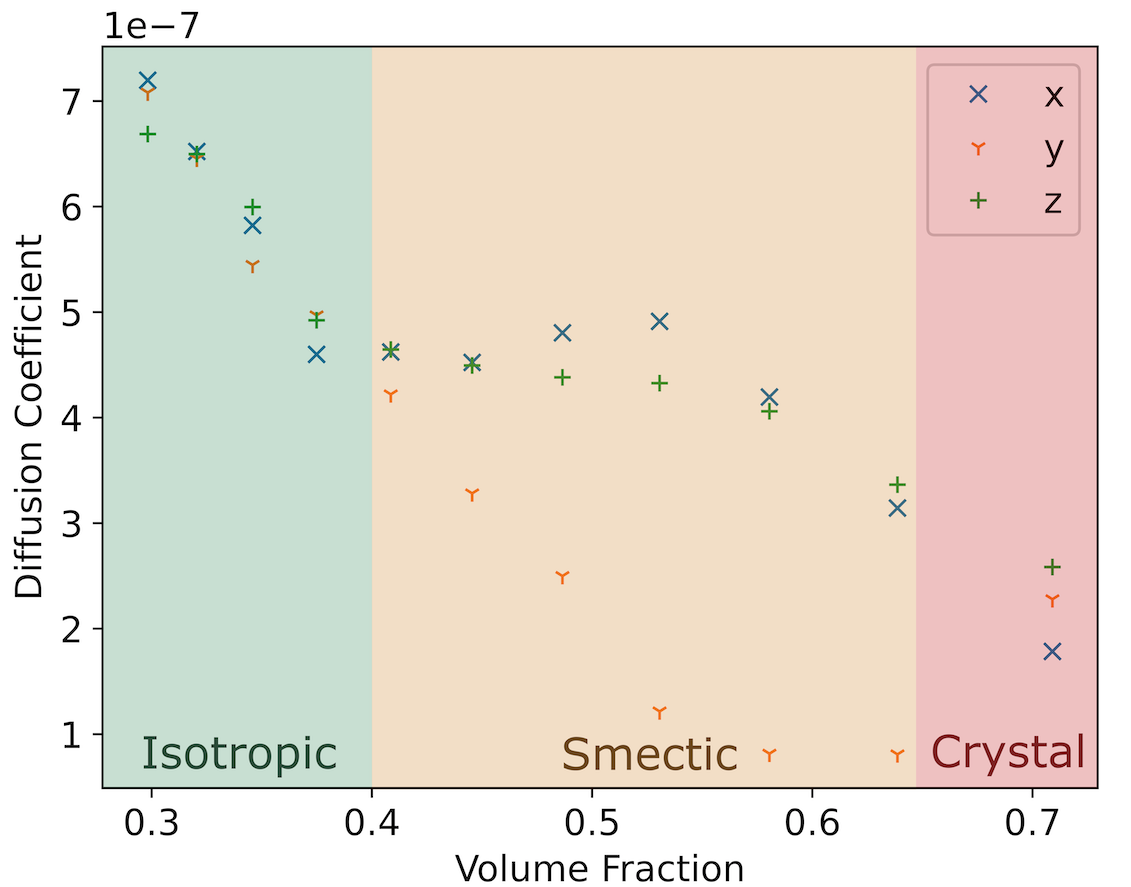}
 \caption{Coordinate diffusion coefficients for the evolution of the microcanonical ensemble at a range of volume fractions $\phi$. The low volume fraction region on the left of the graph corresponds to the isotropic phase, with no variation between coordinate axes. In contrast, the smectic phase in the high volume fraction gives rise to anisotropy in the coordinate diffusion coefficients, with reduced diffusivity perpendicular to the smectic layers in the $y$ axis direction.}
 \label{fgr:nun_Dcoeff}
\end{figure*}

Phase formation may be directly observed through deviations in these coordinate-specific diffusion coefficients. Within the nunchaku system, isotropic diffusion is observed at low volume fractions in Figure \ref{fgr:nun_Dcoeff}, but the $y$-coordinate diffusion coefficient is severely reduced beyond this, with an maximal anisotropy of $D_{y}/D_{x} = 0.19$ at $\phi = 0.58$. This strongly suggests the presence of a smectic phase below $\phi = 0.45$, in agreement with the prediction of $ 0.40<\phi<0.44$ obtained from the smectic order parameter.

\begin{figure*}
 \centering
 \includegraphics[height=7.5cm]{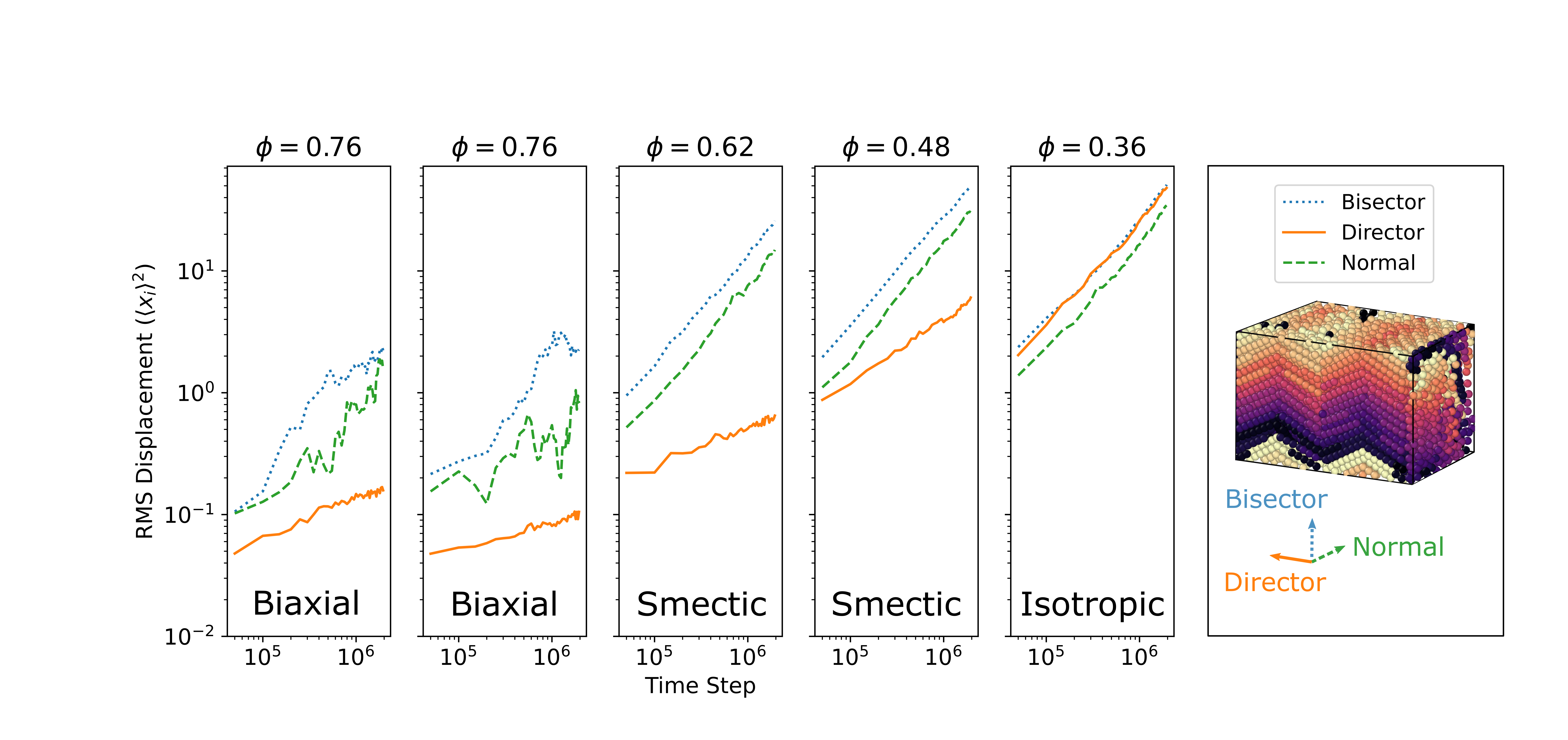}
 \caption{Root mean-squared displacements against time in a system of 1000 bent-core mesogens with an opening angle $120\degree$, over a range of volume fractions $\phi$. As previously, restricted diffusion is observed along the director (previously the $y$-axis) between smectic layers. Anisotropy is now observed within these layer at the highest volume fraction $\phi = 0.76$, suggesting a biaxial smectic phase, with a common aligned bisector. Diffusion is reduced in the direction of the normal vector compared to the bisector vector for these mesogen.}
 \label{fgr:nun_biaxial}
\end{figure*}

Simulation of dynamic properties be utilised to detect novel phases that are not easily identified otherwise. Building on theoretical predictions by Camp et al. \cite{Camp1997, cam991}, we determined that an opening angle $\theta = 120\degree$ would be optimal for biaxial phase formation. We were able to demonstrate the existence of a biaxial smectic phase through the measurements of directional diffusion coefficients, as depicted in Figure \ref{fgr:nun_biaxial}. The system was initially configured as a crystalline lattice of linear molecules, before the minima in the opening angle was shifted to form the desired opening angle. This process occurred randomly, and so no preferential bisector direction was observed. Subsequent equillibration (\num{2e6} steps) allowed relaxation into the smectic biaxial state, which was then observed through repeated simulations (\num{2e6} steps) to transition back into the traditional smectic and isotropic phases upon expansion.

This demonstrates the existence of a biaxial smectic phase for this system of DNA mesogens, and the power of dynamic property simulation in novel phase identification.

\section{Conclusion}

We considered the phase behaviour of DNA nunchaku molecules, consisting of two sections of ds-DNA connected by a short section of ss-DNA. We introduced a coarse-grained model of DNA nunchaku particles, formed of two rigid rods connected by a (semi-) flexible linker, with soft-core, purely repulsive potential interactions. We also introduced a simpler `rigid rod' model, without the flexible linker, to verify the phase identification techniques used here.

Our simulations of rigid rods were consistent with Onsager's prediction of the existence of a first-order `entropic' phase transition between the isotropic and nematic phases of slender hard rods, across a wide range of aspect ratios. In comparison to Onsager's predictions of a transition volume fraction $\phi = 0.4$ for rods with aspect ratio $L/D = 10$, we were able to confine the measured critical volume fraction within the range $0.39<\phi<0.44$. This value, along with the equilibrium nature of this transition, was verified through simulations prepared an initial crystalline phase and subject to a stepwise expansion of the simulation box, eliminating the possibility of phase hysteresis. 

The application of the same techniques to the nunchaku system gave evidence for the formation of an ordered, quasi-nematic phase in this system, in both the semi-flexible ($S_{n} = 0.58$) and bent core ($S_{n} = 0.62$) configurations. The fixed rigidity configuration, a more realistic representation of the DNA system, demonstrated clear evidence for an entropy-driven phase transition, through the reduction in configurational entropy associated with the formation of a preferred angle. The pair-wise orientational correlation function was also used to validate the presence of truly long-range order.

Finally, we demonstrate that the measurement of dynamic properties provide a little-studied, alternative method for identifying phase transitions and characteristic symmetries of such systems, in comparison with phases previously identified in both the rigid-rod and nunchaku systems. In particular, we considered the formation of anisotropic phases through the variation in coordinate diffusion coefficients, such as in the smectic phase where the diffusion coefficient within the ordered layers is up to a factor of $5.15$ greater than between them. We extend this method beyond the simple Cartesian approach taken here, to allow the identification of novel phases where the global director or symmetry class is not known. Using this approach to directional diffusion within the nunchaku system, we also demonstrate the formation of a novel biaxial phase, not previously observed in DNA-based systems. We suggest this method can be used in further identification of novel liquid crystal phases with directional anisotropy.

Through a greater understanding of the phase behaviour of DNA nanoparticles, we may obtain better insight for the design of more complex mesogen shapes, and thus develop new, entropy-driven self-assembled structures \cite{Barry2010, Lin2000}. These techniques may also have applications in biophysics, photonics, structural biology and synthetic biology \cite{Nummelin2018, Praetorius2017, Bathe2017}.


\begin{acknowledgments}
K. G. acknowledges funding from the Streetly Fund for Natural Sciences at Queens' College, Cambridge. J. Y. and R. L. thank the Cambridge Trust and China Scholarship Council (CSC), and D.A.K. thanks the UK Engineering and Physical Sciences Research Council Ph.D. Studentship (award No. 1948692) for financial support. E. E. acknowledges support of the Research Council of Norway through its Centres of Excellence funding scheme, project number RCN 262644.
\end{acknowledgments}

\section*{Data Availability Statement}

The data that support the findings of this study, along with all simulation and analysis code, are openly available at \url{https://github.com/KCGallagher/LC_Project}.

\section*{Conflicts of Interest}

The authors have no conflicts to disclose.

\bibliography{main}

\appendix

\section{Detailed simulation model from oxDNA to LAMMPs}

In this study we utilised the advanced oxDNA2 model\citep{snodin2015introducing}, which reflects the minor and major groove caused by the different sequences of base-pairings AT and CG in the DNA helix. As opposed to the original oxDNA model \citep{doye2013coarse, linak2011erratum} the extended version also considers the effect of salt on the screening of the Coulomb repulsion between the negatively charge backbones mimicking real experimental realisations of DNA mesogens. 

\begin{figure}[htbp!] 
\centering
\includegraphics[width=\columnwidth]{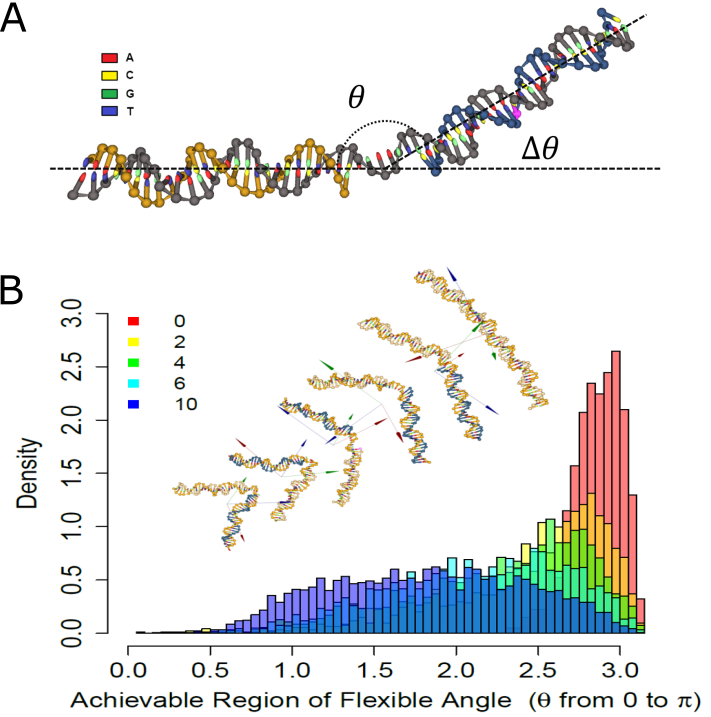}
\caption[F2 hist]{(A) The model of a nunchaku DNA, with $\theta$ being the bending angle, and $\mathrm{\Delta}\theta = \pi -\theta$. (B) Probability of $\cos\theta$ of bending angles  achievable for different number of non-binding nucleotides placed at the centre of a nunchakuDNA; the red bars represent the quasi-rigid dsDNAwith no missing part, while the purple bars represent 18 unbound bases\,\cite{Yu2022}.} \label{fig:SI1} 
\end{figure}

The nunchakus were based on a 60 base-pair long dsDNA forming a semi-rigid rod that served as reference. The detailed DNA sequence is shown in the section below. Because large amounts of DNA are needed to observe liquid crystal phase transitions in experiments, we only tested the thermal stability of the sequences we studied in simulations. This was done to also verify the soundness of our oxDNA simulations. We focused only on nunchaku DNA with short and long flexible joints. We know that dsDNA can be considered as a polymer with some rigidity, hence many currently existing bead-bond DNA models are treating dsDNA as a string of rigidly connected beads \citep{doye2013coarse, linak2011erratum}. The presence of the ssDNA chain between the two dsDNA arms in our nunchaku DNA provides extra flexibility at a specific place along the rod so that two dsDNA can be freely bent. So far we did not find an existing model that can accurately describe this bending property. Another important concern is the different length of the ssDNA joint, e.g. the long flexible chain can be fully bent, but the short flexible chain can only provide a limited bending angle. In order to solve these problems, we use a simpler but relatively accurate method to simulate these nunchaku-DNAs in a dense state, described in the PhD thesis of J. Yu\,\cite{Yu2022}. First we explored a simple potential equation that can describe this bending behaviour. In Figure\,\ref{fig:SI1}, we plotted a statistical measurement in nunchuck-DNAs with different numbers of build-in flexible ssDNA bases. A full analysis of the bending free energy between the rigid dsDNA arms is also provided in the PhD thesis of J. Yu. Here we give the essential results

The bending free energy determines the probability of finding a particular angle between the dsDNA arms of the nunchaku and must be a function of the number of nin-binding ssDNA bases in the joint. 
Hence, the angle $\theta$ between the rigid arms  (Fig.\,\ref{fig:SI1}A) was calculated using two vectors $\vec{r_1}$ and $\vec{r_2}$ defined by the arms by:

\begin{equation}
    \cos\theta = \frac{\vec{r_1}\cdot\vec{r_2}}{ \mid\vec{r_1}\mid\cdot\mid\vec{r_2}\mid}.
\end{equation}

\noindent
Therefore, the probability, and hence the free energy, of finding a particular angle between the arms of the nunchaku depends only on $\cos\theta$. The probability of finding a specific angle between the dsDNA arms $P(\cos \theta)$ can then be written as:
\begin{equation}\label{eq:pcos}
P(\cos\theta) = \dfrac{1}{Z} \cdot e^{-\dfrac{F(\cos\theta)}{k_{B}T}},
\end{equation}

\noindent
where $k_B$ is the Boltzmann constant, and $T$ is the absolute temperature, with $Z$ being the partition function to ensure the probability is normalised. 
A parameter that can quantitatively describe the level of flexibility in each base length can be found by re-writing the Eqn.\,\ref{eq:pcos} as a function of $\cos\theta$:

\begin{equation}\label{eq:fcos}
F(\cos\theta) = -k_{B}T \cdot \ln[Z \cdot P(\cos\theta)] =  -k_{B}T \cdot \ln P(\cos\theta) + C.
\end{equation} 

\noindent
Here $C=k_{B}T \cdot \ln Z$ is a normalised constant (relate to the partition function $Z$). 

\begin{figure}[htbp!] 
\centering
\includegraphics[width=\columnwidth]{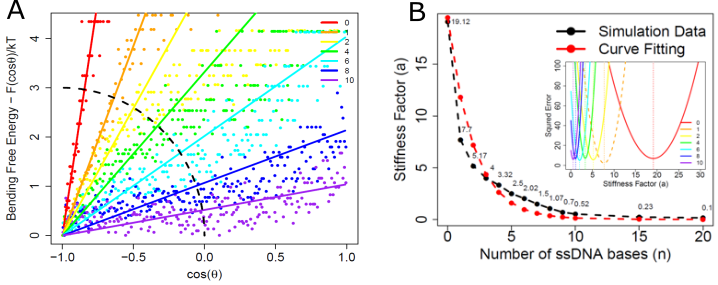}
\caption{ (A) Statistical results of bending free energy $F(\cos\theta)$ calculated from $\cos\theta$ distributions. The squared error shows quantitative data to find the best fitting curve within the black dash region. (B) The stiffness factor $a$ is the gradient slope of the best linear fitting line. The
stiffness factor represents the flexible level according to different lengths of ssDNA joints
used in nunchucks DNA. A smaller value of the stiffness factor means higher flexibility.  in a nunchucks DNA; the numbers 0, 1, 2, 4, 6, 8 and 10 represent the counts of ssDNA bases we used in our flexible chain between two dsDNA arms. Zero (red) refers to the full DNA duplex.} \label{fig:SI2} 
\end{figure}

In order to determine the bending free energy, we need to know the probability distribution of all achievable angle states ($P$). This probability distribution was determined from our simulated outcomes by monitoring the bending angle frame by frame.  
We plotted statistical results of a flexible bending angle distribution in Fig.\,\ref{fig:SI1}, based on specific lengths of ssDNA joints ranging from 0 to 10 bases. All bending angle calculations were taken from the equilibrium state when the overall energy of the nunchaku DNA was measured to be stable in oxDNA simulations. Then we pick up ten individual periods in the equilibrated period to calculate the $\cos\theta$ distribution. Each run provided at least one thousand measured samples to ensure statistical accuracy. For the simulated nunchaku DNA to be fully flexible, we should expect a horizontal line for the $\cos\theta$ distribution. From our statistical data, it was noticeable that flexible joints with eight bases or higher numbers were close to a horizontal line, trending to a fully flexible state. 
To see the relationship more clearly, we plotted the bending free energy ($F(\cos\theta)$) as as function of $\cos\theta$ in Fig.\,\ref{fig:SI2}. This can then be used to fit the parameters in the simplest possible form for $F(\cos \theta)$, which we will determine in the following section.

The statistical data in Fig.\,\ref{fig:SI2} are fitted by linear regression assuming $F(\cos \theta)$ is linear in its argument.  Considering an ideal nunchaku DNA, the equilibrium angle between its two arms should be $\theta_{eq} \approx \pi$ (Fig.\,\ref{fig:SI1}A). At this most relaxed state, the bending free energy is at a minimum. Adding a small fluctuation $\Delta\theta$, to the equilibrium state, with $\Delta\theta<<1$, the bending free energy near the equilibrium state increases slightly, and the bending angle ($\theta$) and the applied fluctuation ($\Delta\theta$) will have the following relation:

\begin{equation}
 \theta = \theta_{eq} - \Delta\theta  = \pi - \Delta\theta.
\end{equation}

\noindent    
The updated bending free energy due to a small angle fluctuation is then 

\begin{equation}
F(\cos(\theta_{eq}-\Delta\theta)) = F(\cos\theta_{eq}\cos\Delta\theta + \sin\theta_{eq}\sin\Delta\theta).
\end{equation}

\noindent
Consider $\theta_{eq} \simeq \pi$, we have $\sin\theta_{eq} = 0$.  The above equation can be rewritten to be:
\begin{equation}
F(\cos(\theta_{eq} - \Delta\theta)) = F(\cos\theta_{eq}\cos\Delta\theta).
\end{equation}

\noindent
Using a Taylor expansion, assuming $\Delta \theta \ll 1$ and using $\cos(\theta_{eq})=\cos\pi =-1$, we get,

\begin{equation}
F(\cos(\theta_{eq} - \Delta\theta)) \simeq F(\dfrac{1}{2}(\Delta\theta)^2-1)= F(\cos\theta_{eq}+\dfrac{1}{2}(\Delta\theta)^2).
\end{equation}

\noindent
And using Taylor's expansion again, we get,
\begin{equation}
 F(\cos\theta_{eq}+\dfrac{1}{2}(\Delta \theta)^2) \simeq F(\cos\theta_{eq})+\dfrac{1}{2} (\Delta \theta) ^2 F'(\cos\theta_{eq}) + O((\Delta \theta)^4),
\end{equation}

\noindent
with $O((\Delta\theta) ^4) \simeq 0$. We may define the free energy at equilibrium to be zero. Again from the Taylor expansion, we have $\dfrac{1}{2} (\Delta \theta) ^2 = 1-\cos\Delta\theta$ and from the angle definition $\Delta \theta = \pi -\theta$, we have $(1-\cos\Delta\theta) = (1-\cos(\pi-\theta))=(1+\cos\theta)$. Hence, the above equation can be rewritten as:

\begin{equation} \label{eq:angle39}
F(\cos(\theta_{eq} - \Delta\theta)) \simeq (1+\cos\theta) F'(\cos\theta_{eq}).
\end{equation}

For non-ideal nunchucks DNA, the $\theta_{eq}$ might not be perfectly equal to $\pi$. It has a small bias because of the asymmetry due to the break in the backbone of one of the ssDNA strands, but it should not affect this calculation. In 3D, we assume the actual fluctuations of the angle can be in any direction, but no matter which direction, we always have $\theta = \theta_{eq} -\Delta\theta$, as the nunchucks-model considered to be symmetric. In summary, the simplest prediction of a linear relation can be written as

\begin{equation}\label{eq:fcos_final}
F(\cos(\theta)) = \alpha_{n} \epsilon (1+\cos\theta),
\end{equation}

\noindent
where $\epsilon$ has units of energy, $\alpha_{n}$ is the stiffness factor $a$ to represent the flexibility of the ssDNA joints. It is simply the gradient slope of the bending free energy from Eqn.\,\ref{eq:angle39}. $\alpha_{n}$ is a discrete function ($f(n)$) that depends on the number of non-binding ssDNA bases ($n$) used:

\begin{equation}
\alpha_{n} = f(n).
\end{equation}

In Table\,\ref{tab:fn} we list the stiffness factors for different $n$ obtained from evaluations of long runs of the oxDNA-simulations: 

\begin{table}[htbp!]
\begin{tabular}{c|c|c|c|c|c|c|c|c|c|c|c|c|c}
 $n$  & 0     & 1    & 2    & 3    & 4    & 5    & 6    & 7    & 8    & 9    & 10   & 15   & 20      \\ \hline
$f(n)$ & 19.12 & 7.7 & 5.17 & 4.0 & 3.32 & 2.5 & 2.02 & 1.5 & 1.07 & 0.7 & 0.52 & 0.23 & 0.1   \\ 
\end{tabular}
\caption{Values of stiffness factor f(n) according to the number of bases on the ssDNA flexible joint in nunchucks DNA. The data were extracted from the statistical analysis in Fig.\,\ref{fig:SI2}\,\cite{Yu2022}.}\label{tab:fn}
\end{table}

\subsection{From oxDNA to a second-level coarse-grained simulation model}

\begin{figure}[htbp!] 
    \centering
    \includegraphics[width=\columnwidth]{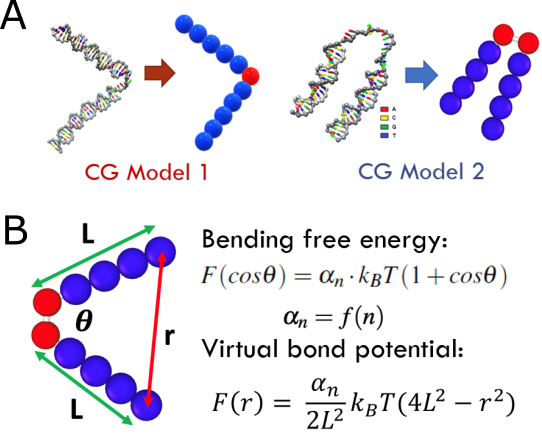}
      \caption{(A) Coarse-raining model design. CG model 1 fits nunchucks DNA with short chain length ($N_s <6$); CG model 2 works for longer flexible linkers ($N_s > 6$). (B) Translation of the bending free energy into a virtual bond potential in CG 2. $L$ is the arm length, $\theta$ is the bending angle, $r$ is the virtual distance between the head of the two arms and $\alpha_n$ is the stiffness factor\,\cite{Yu2022}.} \label{fig:IS3} 
    \end{figure}

We considered two types of coarse-grained (CG) models when designing the second-level coarse-grained model. The first model [CG Mode 1] has one single flexible bead in the middle, which fits the nunchaku DNA with a short flexible joint when it is within a confined region ($N_s < 6$). The second model [CG Mode 2] has two flexible beads in the middle, which fits the nunchaku DNA with a longe flexible joint ($N_s > 6$) Fig.\,\ref{fig:IS3}).

The other concern of designing CG model types according to their joint length is the flexible saturation we observed in the `fully flexible' nunchucks DNA. Once the model gets the flexible region, the cost of bending is almost negligible, i.e. the stiffness factor is nearly equal to zero. Hence, in [CG model 2], we can release the bending confinement and let the arms be fully flexible to move. While for the model design for liquid crystal study, we suggest using [CG model 1] in simulating mesogens with confined angles, as it matches better with preferring to bend into a `herringbone' structure.

\section{DNA sequences in oxDNA simulation and experiments}

\begin{verbatim}
F0 (Full DNA): [60//60]
5’ GAG GTG GAG TAA GTG AAT GAA GGT GTG TGA AGA CAA ACA GAG AGG AGA GAA AGA 
GAG TTA 3’   
3’ CTC CAC CTC ATT CAC TTA CTT CCA CAC ACT TCT GTT TGT CTC TCC TCT CTT TCT 
CTC AAT 5’

N1S (Nunchaku DNA with short flexible joint -5 bases joint ): [60//27+28]
5’ GAG GTG GAG TAA GTG AAT GAA GGT GTG TGA AGA CAA ACA GAG AGG AGA GAA AGA 
GAG TTA 3’   
3’ CTC CAC CTC ATT CAC TTA CTT CCA CAC ----- T GTT TGT CTC TCC TCT CTT TCT 
CTC AAT 5’   

N2L (Nunchaku DNA with long flexible joint -18 bases joint ): [60//21+21]
5’ GAG GTG GAG TAA GTG AAT GAA GGT GTG TGA AGA CAA ACA GAG AGG AGA GAA AGA 
GAG TTA 3’  
3’ CTC CAC CTC ATT CAC TTA CTT     -----     -----     CTC TCC TCT CTT TCT 
CTC AAT 5’     

\end{verbatim}

\section{Nematic Order Parameter} \label{sec:NematicOrderAppendix}
Here we outline the motivation and calculation of the nematic order parameter used throughout this report, based on the work of Eppenga and Frenkel \cite{eppenga1984platelets, frenkel1982disks}. Conventionally, the average second Legendre polynomial $\langle P_{2l}(\cos(\theta)) \rangle$ is used to characterise the nematic order of a system. Averaging over a population of $N$ molecules, we may write the nematic order parameter $S_{n}$ explicitly as:

\begin{equation}
S_{n} = \frac{1}{N} \sum_{i=1}^{N} \left( \frac{3}{2} \cos^{2}(\theta_{i})-\frac{1}{2} \right) 
\end{equation}

This expression however relies on knowledge of the system-wide nematic director (i.e. the axis of symmetry of the cylindrical phase), to define $\theta_{i}$. This is not always possible in physical systems where such a unique direction is not externally imposed.

Instead, as detailed by Frenkel et al. \cite{frenkel1985orientational}, we maximise the expression:

\begin{equation} \label{eq:FrenkelNemOrder}
S^{\prime}_{n}(\boldsymbol{\hat{n}^{\prime}}) = \frac{1}{N} \left[ \sum_{i=1}^{N} \left( \frac{3}{2} (\boldsymbol{\hat{n}^{\prime}} \cdot \boldsymbol{\hat{u}_{i}})^{2}-\frac{1}{2} \right) \right]
\end{equation} where $\hat{u_{i}}$ denotes the orientation of the individual molecular axes in the laboratory frame, and $\boldsymbol{\hat{n}^{\prime}}$ is the direction of common alignment, known as the director. In the absence of an electric field, this direction is arbitrary, and determined in practice by infinitesimal perturbations to the system though spontaneous symmetry breaking \cite{forster2018symmetry}. (\ref{eq:FrenkelNemOrder}) may be further simplified to:

\begin{equation}
S^{\prime}_{n} = \frac{1}{N} \left\langle \boldsymbol{\hat{n}^{\prime}} \cdot \textbf{Q} \cdot \boldsymbol{\hat{n}^{\prime}}  \right\rangle, \qquad where \enspace \textbf{Q}_{i} = \frac{3}{2} \boldsymbol{\hat{u}_{i}}\boldsymbol{\hat{u}_{i}}-\frac{1}{2}\textbf{I}
\end{equation}

The tensor order parameter $\langle \textbf{Q} \rangle$ is a traceless symmetric 2nd-rank tensor, with three eigenvalues $\lambda_{+}, \lambda_{0}, \lambda_{-}$ \cite{eppenga1984platelets}. We typically take the largest eigenvalue $(\lambda_{+})$ as the nematic order parameter, a good approximation in limit of large $N$.
In practice, we calculate the eigenvalues of the related tensor $\textbf{M}$:

\begin{equation}
\textbf{M} =  \frac{1}{N} \sum_{i=1}^{N} \boldsymbol{\hat{u}_{i}}\boldsymbol{\hat{u}_{j}}
\end{equation} as this shares eigenvectors with $\textbf{Q}$, and has eigenvectors $\mu_{n}$ related to $\lambda_{n}$ by: $\mu_{n} = 2/3 \lambda_{n} + 1/3$.

It is worth noting that $\lambda_{+}$ is bound above zero, and so does not reach zero in the isotropic phase as would be expected. It is common to use $S =  -2\lambda_{0}$ when considering such disordered systems, as this fluctuates around an average much closer to zero \cite{mountain1977maiersaupe}. We have not followed this usage in the results presented here, to give continuity in the order parameter over the transition (wherein lies the focus of this paper), however this has resulted in an average order parameter in the isotropic phase slightly above zero.

\subsection{Position Dependant Order Parameter} \label{sec:PairWise_Theory}
The pairwise orientational correlation function provides a position-dependent equivalent to the nematic order parameter. For use in simulation, we write this for $N$ particles in the form:

\begin{equation}
g_{l}(r) = \frac{\sum_{i=1}^{N} \sum_{j \neq i} P_{l}(\boldsymbol{\hat{u}_{i}}\cdot \boldsymbol{\hat{u}_{j}}) \Delta(r_{ij}-r)}{\sum_{i=1}^{N} \sum_{j \neq i} \Delta(r_{ij}-r)}
\end{equation} where $\Delta(r_{ij}-r) = 1$ if $r_{ij}$ is inside a spherical shell between $r$ and $r+\delta$, and $0$ otherwise. The thickness of $\delta$ is chosen to maximise the resolution of the function, as too large a value will `wash out' important features, while still maintaining a reasonable sample size. This method is computationally expensive however, as the number of times $P_{L}$ must be calculated scales as $\mathcal{O}(N^{2})$. Furthermore, the storage of all necessary angles is prohibitively expensive (even on modern computers), and limits the resolution possible \cite{soper1998expense}. We may, however, avoid this problem by expanding the function in terms of spherical harmonics, and sum the contributions for all pairs to a given molecule \cite{soper1994harmonics}. This is most easily achieved through application of the spherical harmonics addition theorem:

\begin{equation}
P_{l}(\cos \theta_{ij}) = \frac{4\pi}{2l+1} \sum_{m=-l}^{l} \mathcal{Y}_{lm}(\boldsymbol{\hat{u}_{i}}) \mathcal{Y}_{lm}^{*}(\boldsymbol{\hat{u}_{j}})
\end{equation}

Using this, recognising that $\boldsymbol{\hat{u}_{i}}\cdot \boldsymbol{\hat{u}_{j}} = \cos \theta_{ij}$ and writing  $\Delta(r_{ij}-r) = \Delta_{r}$ for simplicity, we may write the contribution from all particles around particle $i$ as:

\begin{equation}
\sum_{i \neq j} P_{l}(\cos \theta_{ij}) \Delta_{r} =  \frac{4\pi}{2l+1} \sum_{j \neq i} \sum_{m=-l}^{l} \mathcal{Y}_{lm}(\boldsymbol{\hat{u}_{i}}) \mathcal{Y}_{lm}^{*}(\boldsymbol{\hat{u}_{j}}) \Delta_{r}
\end{equation}

Exchanging the order of summation, and removing $\mathcal{Y}_{lm}(\boldsymbol{\hat{u}_{i}})$ from the sum over $j$ gives:

\begin{equation}
\sum_{i \neq j} P_{l}(\cos \theta_{ij}) \Delta_{r} =  \frac{4\pi}{2l+1} \sum_{m=-l}^{l} \mathcal{Y}_{lm}(\boldsymbol{\hat{u}_{i}}) \sum_{j \neq i}  \mathcal{Y}_{lm}^{*}(\boldsymbol{\hat{u}_{j}}) \Delta_{r}
\end{equation}

This may be simplified further to:

\begin{equation} \label{eq:pairwise_finalres}
\sum_{i \neq j} P_{l}(\cos \theta_{ij}) \Delta_{r} =  \frac{4\pi}{2l+1} \sum_{m=-l}^{l} \mathcal{Y}_{lm}(\boldsymbol{\hat{u}_{i}}) \times C_{lm}^{*}
\end{equation} where we have defined

\begin{equation}
 C_{lm}^{*} = \sum_{j \neq i}  \mathcal{Y}_{lm}^{*}(\boldsymbol{\hat{u}_{j}}) \Delta_{r}
\end{equation}

This removes the requirement to calculate $P_{l}$ for every pair of molecules, in favour of simply summing the spherical harmonics terms corresponding to a given shell around $i$, then multiplying by the spherical harmonic components. It should also be noted that (\ref{eq:pairwise_finalres}) is invariant under changes in the coordinate frame; in our computations all orientations were expressed in the lab frame for each, rather than aligning the coordinate frame with the global director.

\end{document}